\documentclass[useAMS,usenatbib,twocolumn]{mn2e}
\usepackage{graphicx}
\usepackage{subfigure}

\newcommand{\mpt}{\mathrm{.}}
\newcommand{\mcm}{\mathrm{,}}

\newcommand{\Vec}[1]{ \mbox{\boldmath$ #1 $} }

\newcommand{\apjl}{ApJ}
\newcommand{\apj}{ApJ}

\newcommand{\mnras}{MNRAS}

\newcommand{\aj}{AJ}
\newcommand{\aap}{A\&A}

\title[A generalised mass-sheet degeneracy]{A generalisation 
of the mass-sheet degeneracy producing ring-like artefacts in 
the lens mass distribution}
\author[J. Liesenborgs, S. De Rijcke, H. Dejonghe and P. Bekaert]
{J. Liesenborgs$^1$\thanks{Corresponding author:
jori.liesenborgs@uhasselt.be}, S. De Rijcke$^2$\thanks{Postdoctoral
Fellow of the Fund for Scientific Research - Flanders
(Belgium)(F.W.O)}, H. Dejonghe$^2$ and P. Bekaert$^1$\\ $^1$
Expertisecentrum voor Digitale Media, Universiteit Hasselt,
Wetenschapspark 2, B-3590, Diepenbeek, Belgium \\ $^2$ Sterrenkundig
Observatorium, Universiteit Gent, Krijgslaan 281, S9, B-9000, Gent,
Belgium}

\begin{document}

\date{} 

\pagerange{\pageref{firstpage}--\pageref{lastpage}} \pubyear{2008}

\maketitle \label{firstpage} 

\begin{abstract} 
The inversion of a gravitational lens system is, as is well known,
plagued by the so-called mass-sheet degeneracy: one can always
rescale the density distribution of the lens and add a
constant-density mass-sheet such that the, also properly rescaled,
source plane is projected onto the same observed images. For strong
lensing systems, it is often claimed that this degeneracy is broken as
soon as two or more sources at different redshifts are available. This
is definitely true in the strict sense that it is then impossible to
add a constant-density mass-sheet to the rescaled density of the lens
without affecting the resulting images. However, often one can easily
construct a more general mass distribution -- instead of a
constant-density sheet of mass -- which gives rise to the same
effect: a uniform scaling of the sources involved without affecting
the observed images. We show that this can be achieved by adding one
or more circularly symmetric mass distributions, each with its own
center of symmetry, to the rescaled mass distribution of the
original lens. As it uses circularly symmetric distributions, this 
procedure can lead to the introduction of ring shaped features in 
the mass distribution of the lens. In this paper, we show explicitly 
how degenerate inversions for a given strong lensing system can 
be constructed. It then becomes clear that many constraints are 
needed to effectively break this degeneracy.
\end{abstract}


\section{Introduction}

Being an ill-posed problem, it is not surprising that gravitational 
lens inversion is plagued by degeneracies. Several classes of 
degeneracies were first identified by \citet{Gorenstein} and were
later reinterpreted by \citet{Saha2000} in terms of changes in the
arrival-time surface. Of these, the most widely known
degeneracy is usually called the mass-sheet degeneracy, although
recently the more correct name of steepness degeneracy has been
suggested \citep{SahaSteepness}. This degeneracy was first mentioned
in the context of the strong lensing system Q0957+561 
\citep{FalcoMassSheet}, but has also been studied in weak lensing
systems (e.g. \citet{SchneiderMassSheet}) and even in the context
of microlensing \citep{Paczynski}.

In strong lensing systems, it is often claimed that the presence of
two sources at different redshifts suffices to break the mass-sheet
degeneracy (e.g. \cite{Abdelsalam2}). While it is definitely 
true that a constant-density
mass-sheet can no longer be used to construct degenerate solutions in
such a case, we show in this paper how a more general mass
distribution, to be added to the original lens mass distribution, can
be constructed which gives rise to a similar type of degenerate
solutions. In particular, the alternative name of steepness degeneracy
is still applicable since the construction of degenerate solutions
still requires the original mass distribution to be rescaled. Although
the method described below is quite straightforward, the authors,
having performed a thorough literature study, are not aware of this
result being published before.

In section \ref{sec:formalism} the necessary equations from the
gravitational lensing formalism will be reviewed. These equations
will be used to derive the mass-sheet degeneracy in section
\ref{sec:masssheet} and to explain the extension to multiple
redshifts in section \ref{sec:extension}. Finally, in 
section \ref{sec:conc}, the results and their implications are 
discussed.

\section{Lensing formalism}\label{sec:formalism}

We shall briefly review the necessary equations related to the
gravitational lensing formalism. The interested reader is referred to
\citet{SchneiderBook} for a thorough review of gravitational lensing
theory.

In the thin lens approximation, the gravitational lens effect
is essentially a mapping of the source plane ($\beta$-space)
onto the image plane ($\theta$-space), described by the lens equation:
\begin{equation}
\Vec{\beta}(\Vec{\theta}) = \Vec{\theta} - 
\frac{D_{\rm ds}}{D_{\rm s}}\Vec{\hat{\alpha}}(\Vec{\theta})\mpt
\end{equation}
Here, $\Vec{\hat{\alpha}}(\Vec{\theta})$ describes the instantaneous
deflection of a light ray that punctures the lens plane at the
position $\Vec{\theta}$, and depends linearly on the two-dimensional
projected surface mass density $\Sigma(\Vec{\theta})$ of the lens.

For a circularly symmetric projected mass density centered on the 
origin, the defection angle reduces to:
\begin{equation}\label{eq:alphasymm}
\Vec{\hat{\alpha}}(\Vec{\theta}) = \frac{4 G M(\theta)}
{c^2 D_{\rm d} \theta^2}\Vec{\theta}\mcm
\end{equation}
in which $M(\theta)$ is the total mass within a radius $\theta$.  
The geometry of the situation is described by the angular diameter
distances $D_{\rm d}$, $D_{\rm ds}$ and $D_{\rm s}$. As a
result, for a circularly symmetric projected mass distribution, only
the total mass inside a specific radius contributes to the deflection
of light rays.

\begin{figure*}
\centering
\subfigure{\includegraphics[width=0.44\textwidth]{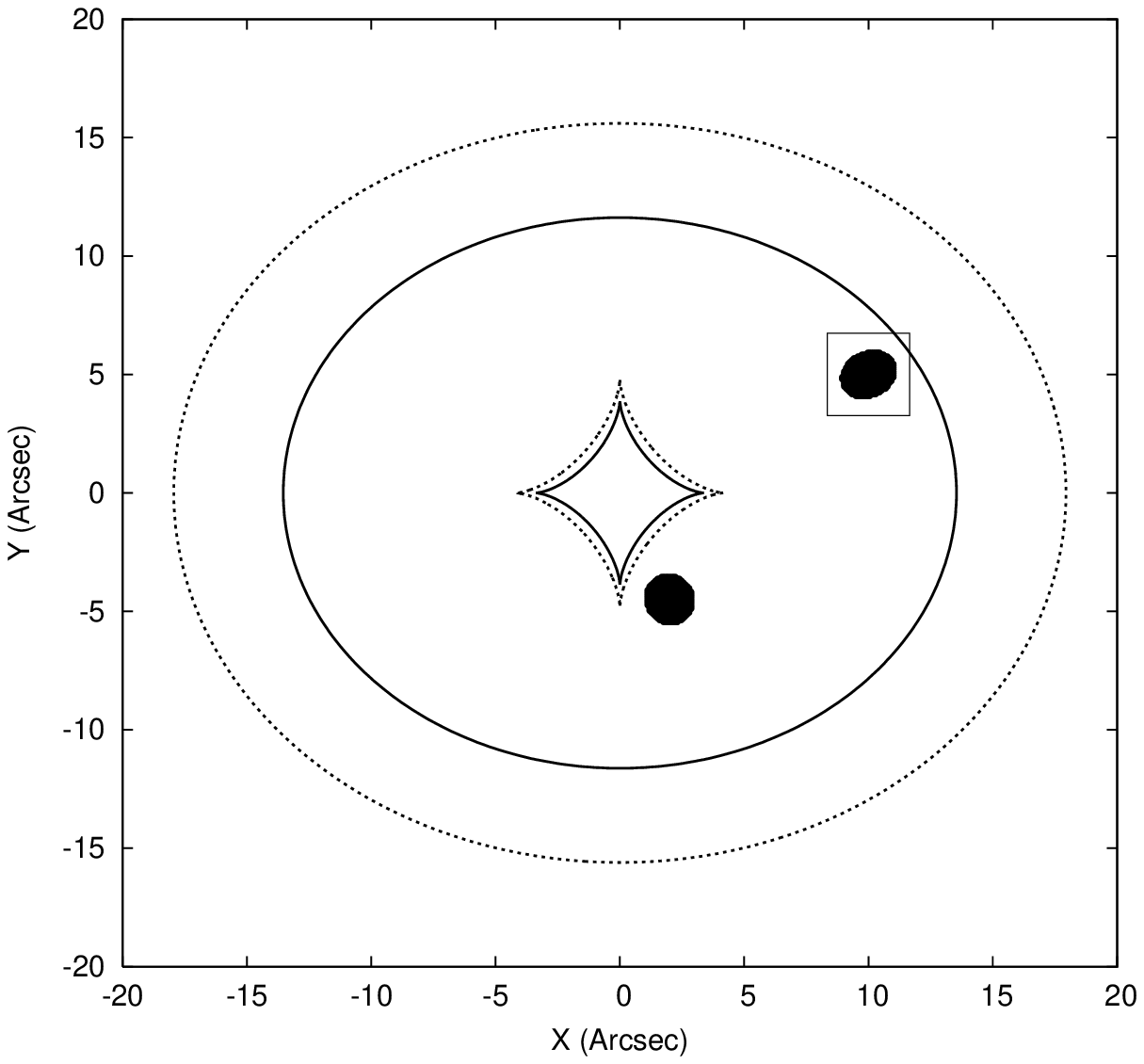}}
\qquad
\subfigure{\includegraphics[width=0.44\textwidth]{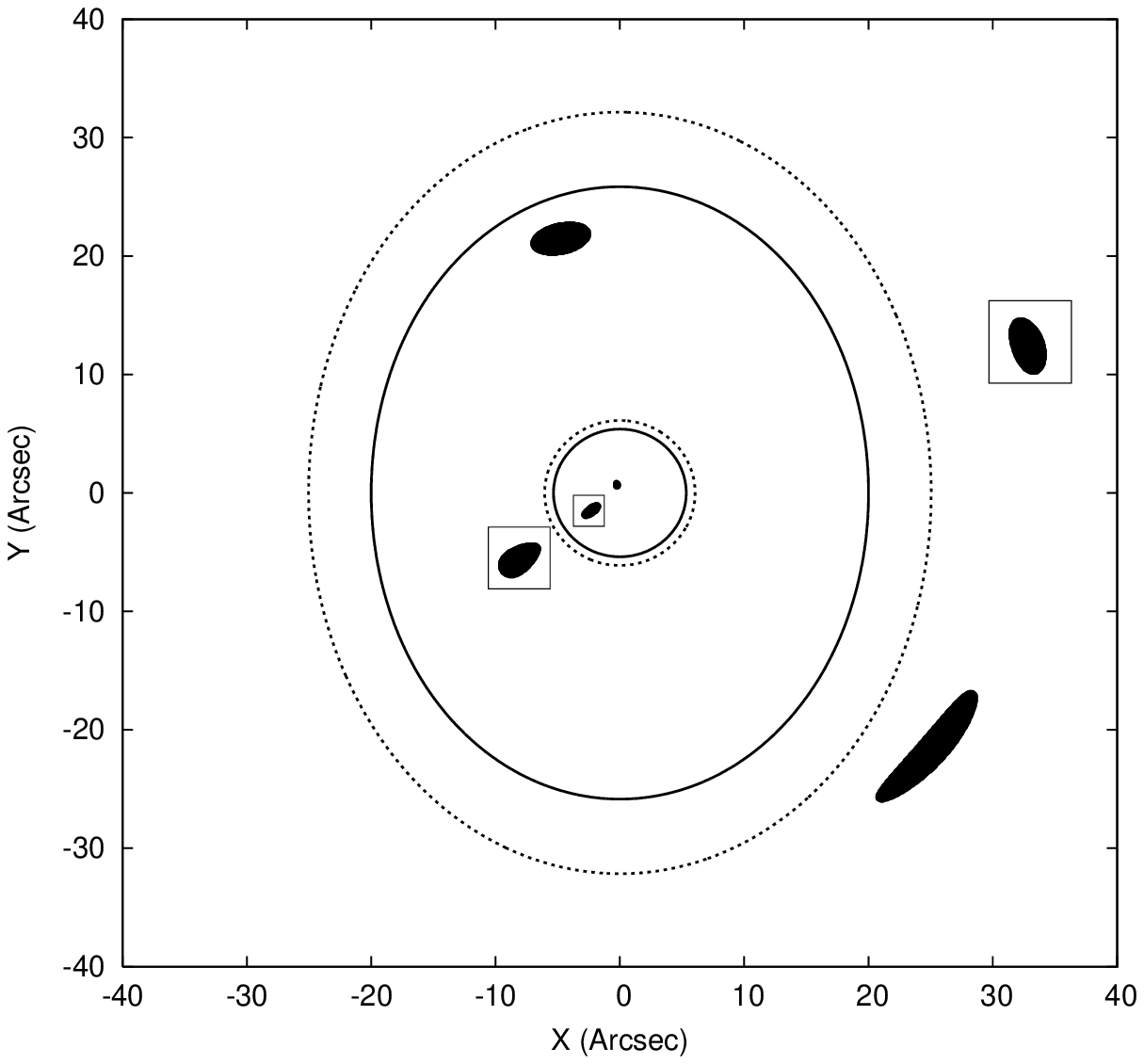}}
\caption{Left panel:~original sources used to illustrate the construction
of degenerate solutions. The source surrounded by a box is placed
at redshift $z_1=1.2$, the second source is at $z_2=1.8$. The caustics
created by the non-singular isothermal ellipse placed at $z=0.5$ are
also visible. The solid line corresponds to $z_1=1.2$, the dotted line
to $z_2=1.8$. Right panel:~images of the two sources used to illustrate the 
construction of degenerate solutions. The critical lines are also shown.}
\label{fig:planes}
\end{figure*}

\begin{figure*}
\centering
\subfigure{\includegraphics[width=0.44\textwidth]{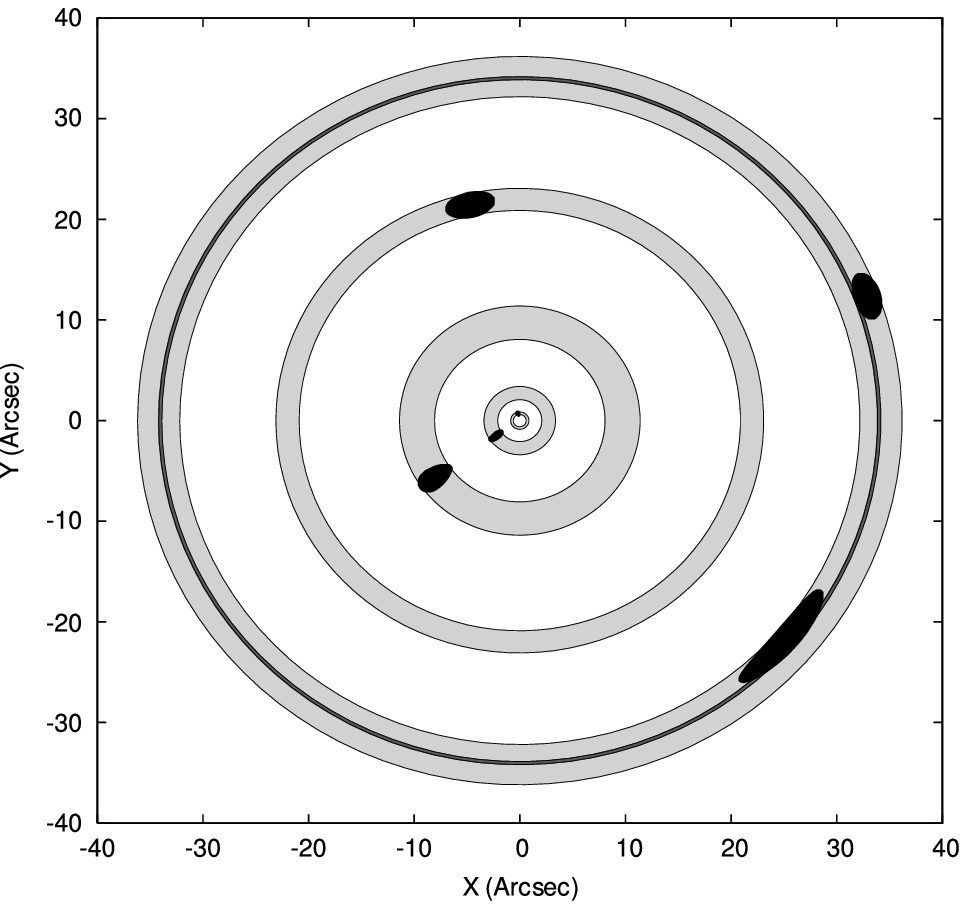}}
\qquad
\subfigure{\includegraphics[width=0.44\textwidth]{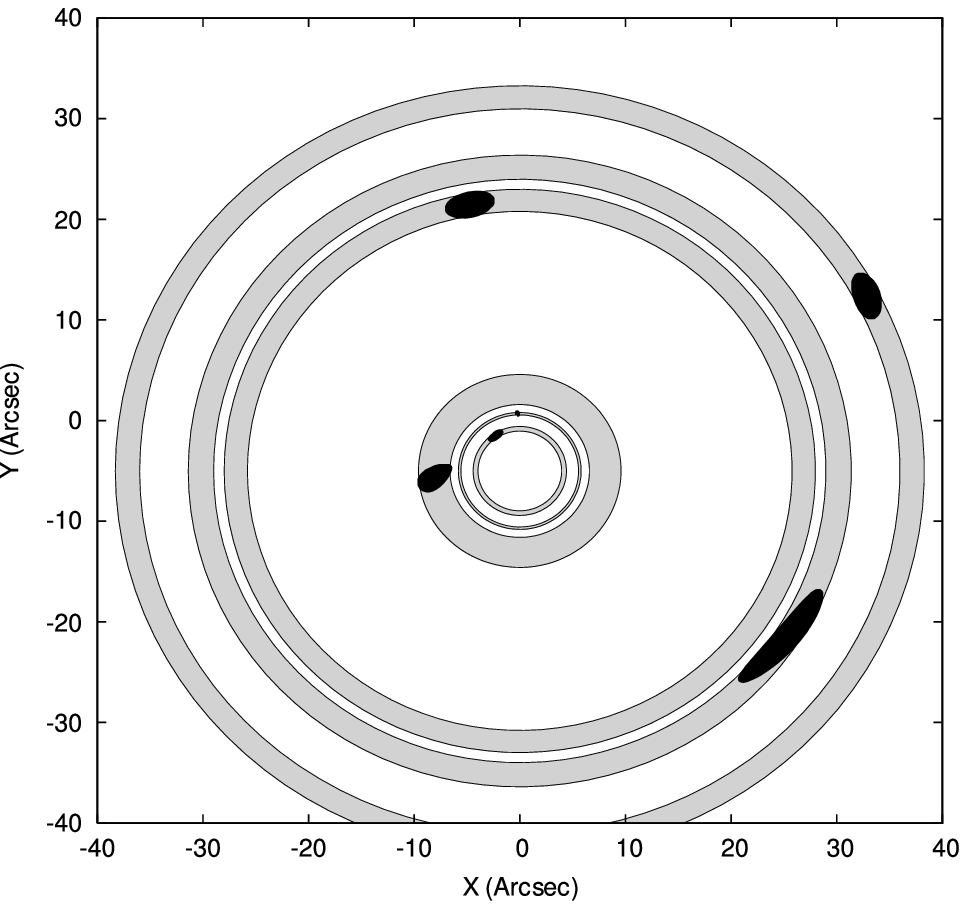}}
\caption{Left panel:~the annuli in which the images reside, as seen
from the origin of the coordinate system, are
displayed as grey rings. The darker ring indicates the region
in which the annuli of the outer images overlap. Because of this overlap,
no suitable mass density can be constructed (see text).
Right panel:~similar to the left panel, but now the annuli are
centered on $(0, -5)$. This center can be used in the construction
of a degenerate solution since there are no longer overlapping
annuli.}
\label{fig:imagepositions}
\end{figure*}

\section{The mass-sheet degeneracy}\label{sec:masssheet}

Let us first consider a strong lensing system with images coming from
a single source. A uniform sheet of mass with density $\Sigma_{\rm s}$
produces a deflection described by
\begin{equation}\label{eq:sigma0}
\Vec{\hat{\alpha}}_{\rm s}(\Vec{\theta}) = 
\frac{D_{\rm s}}{D_{\rm ds}}\frac{\Sigma_{\rm s}}{\Sigma_{\rm cr}}\Vec{\theta}\mcm
\end{equation}
in which the critical mass density for the current geometry is defined as
follows:
\begin{equation}
\Sigma_{\rm cr} = \frac{c^2}{4 \pi G D_{\rm d}}\frac{D_{\rm s}}{D_{\rm ds}}\mpt
\end{equation}
Note that $\Sigma_{\rm cr}$ depends on the redshift of the source via the
angular diameter distances $D_{\rm s}$ and $D_{\rm ds}$. Let
$\Sigma_0(\Vec{\theta})$ be a mass distribution that is compatible
with the observed images. This means that the corresponding lens
equation
\begin{equation}
\Vec{\beta}_0(\Vec{\theta}) = \Vec{\theta} - 
\frac{D_{\rm ds}}{D_{\rm s}}\Vec{\hat{\alpha}}_0(\Vec{\theta})
\end{equation}
projects the images onto the source plane in such a way that they
overlap exactly. Without further
constraints, this immediately yields an infinite number of alternative
solutions. Indeed, if the mass distribution
is replaced by
\begin{equation}\label{eq:sheetdegen}
\Sigma_1(\Vec{\theta}) = 
\lambda\Sigma_0(\Vec{\theta}) + (1-\lambda)\Sigma_{\rm cr}\mcm
\end{equation}
the new lens equation becomes
\begin{equation}
\Vec{\beta}_1(\Vec{\theta}) = \Vec{\theta} - 
\lambda \frac{D_{\rm ds}}{D_{\rm s}}\Vec{\hat{\alpha}}_0(\Vec{\theta})
- (1-\lambda) \frac{D_{\rm ds}}{D_{\rm s}}\Vec{\hat{\alpha}}_{\rm s}(\Vec{\theta})
 = \lambda \Vec{\beta}_0(\Vec{\theta})\mpt
\end{equation}
The transformation (\ref{eq:sheetdegen}) describes the so-called 
mass-sheet degeneracy and simply rescales the source plane by 
the factor $\lambda$, producing an equally acceptable source
reconstruction. Note that merely adding a mass-sheet is not
sufficient; one also needs to rescale the original mass distribution
by the same factor $\lambda$, which justifies the alternative name
of steepness degeneracy. The density of the mass-sheet has to be precisely
the critical mass density for this to work. For this reason,
a mass-sheet cannot be used when there are sources at different
redshifts, since these would require different critical densities.

\section{Extension to multiple redshifts}\label{sec:extension}

An infinite sheet of mass, however, is not the only mass distribution
which can be used to produce degenerate solutions. If the mass density
is circularly symmetric and equal to $\Sigma_{\rm cr}$ in an area large
enough to encompass all the images, the same source scaling effect
will arise, thanks to equation (\ref{eq:alphasymm}). The center of
symmetry of such a distribution determines the center of the scaling
(which is the origin of the coordinate system in case of an infinite
sheet). This way, the mass-sheet degeneracy is easily transformed into
a mass-disk degeneracy. In fact, the added mass density need not be
constant inside such a disk to produce the same effect. As long as the
total mass inside each image point is the same as for the mass-disk,
equation (\ref{eq:alphasymm}) ensures that the distribution can be
used to construct a degenerate solution as well. This constraint 
automatically implies a density equal to $\Sigma_{\rm cr}$ inside the 
annuli in which the images reside, but otherwise allows a lot of 
freedom.

This freedom allows us to construct a mass distribution which, when
added to a scaled version of an existing solution for the lens mass
density, is equally compatible with the observed images, but which
will rescale the sources. The effect is therefore very similar to that
of the mass-sheet degeneracy, but this degeneracy is not necessarily 
broken by the presence of additional images of sources at different 
redshifts.

To illustrate the procedure, consider the two sources and their
respective images in Fig. \ref{fig:planes}. The two sources are placed
at redshifts $z_1=1.2$ and $z_2=1.8$ and the images are created by a
non-singular isothermal ellipse at $z=0.5$. This non-singular
isothermal ellipse then provides us with the initial mass density
$\Sigma_0(\Vec{\theta})$. A flat cosmological model with $\Omega_{\rm m} =
0.27$, $\Omega_\Lambda = 0.73$ and $H_0 =
70\,\textrm{km}\,\textrm{s}^{-1}\,\textrm{Mpc}^{-1}$ was used to
calculate the necessary angular diameter distances.

The circularly symmetric mass density $\Sigma_{\rm gen}(\theta)$
and corresponding $M_{\rm gen}(\theta)$ that we shall construct, must 
have the same effect as a mass-sheet for both sources.
This mass density will serve as the generator of the transformation
which creates a degenerate solution $\Sigma_1(\Vec{\theta})$ from an 
existing solution $\Sigma_0(\Vec{\theta})$. The procedure is very
similar to the mass-sheet case:
\begin{equation}\label{eq:addsig}
\Sigma_1(\Vec{\theta}) = \lambda \Sigma_0(\Vec{\theta}) + 
(1-\lambda) \Sigma_{\rm gen}(|\Vec{\theta}-\Vec{\theta}_{\rm c}|)\mcm 
\end{equation}
in which $\Vec{\theta}_{\rm c}$ is the center of symmetry of the generator.
The mass distribution of the generator must satisfy constraints provided
by the images: the mass enclosed by each image point must equal the
mass of the corresponding constant-density mass-sheet. Therefore,
if a specific image of a source at redshift $z$ lies in an annulus
with inner radius $\theta_{\rm in}$ and outer radius $\theta_{\rm out}$,
the constraint provided by said image is the following:
\begin{equation}\label{eq:con}
\forall \theta \in [\theta_{\rm in},\theta_{\rm out}]: M_{\rm gen}(\theta) = \pi D_{\rm d}^2\theta^2\Sigma_{\rm cr}(z)\mcm
\end{equation}
in which the radii are measured with respect to the chosen center
of symmetry $\Vec{\theta}_{\rm c}$.
Consequently, within such an annulus the mass density must equal the
critical density for an image at redshift $z$ and in the region enclosed
by the annulus, the mean density must equal the critical density.
In the left panel of Fig. \ref{fig:imagepositions}, we plot the annuli
of the images, as seen from the center of the non-singular
isothermal ellipse. Looking at the furthest image of each source, it
is clear that no $\Sigma_{\rm gen}$ can be constructed. The mass density
would have to be equal to $\Sigma_{\rm cr}(z_1)$ inside the annulus of one
image and $\Sigma_{\rm cr}(z_2)$ inside the annulus of the other
image. Since these regions overlap, as is indicated by the darker ring,
this is impossible. However, if we take $(0, -5)$ as the center, 
there are no overlapping annuli as can be seen in the right panel of 
Fig. \ref{fig:imagepositions}.

\begin{figure*}
\centering
\subfigure{\includegraphics[width=0.44\textwidth]{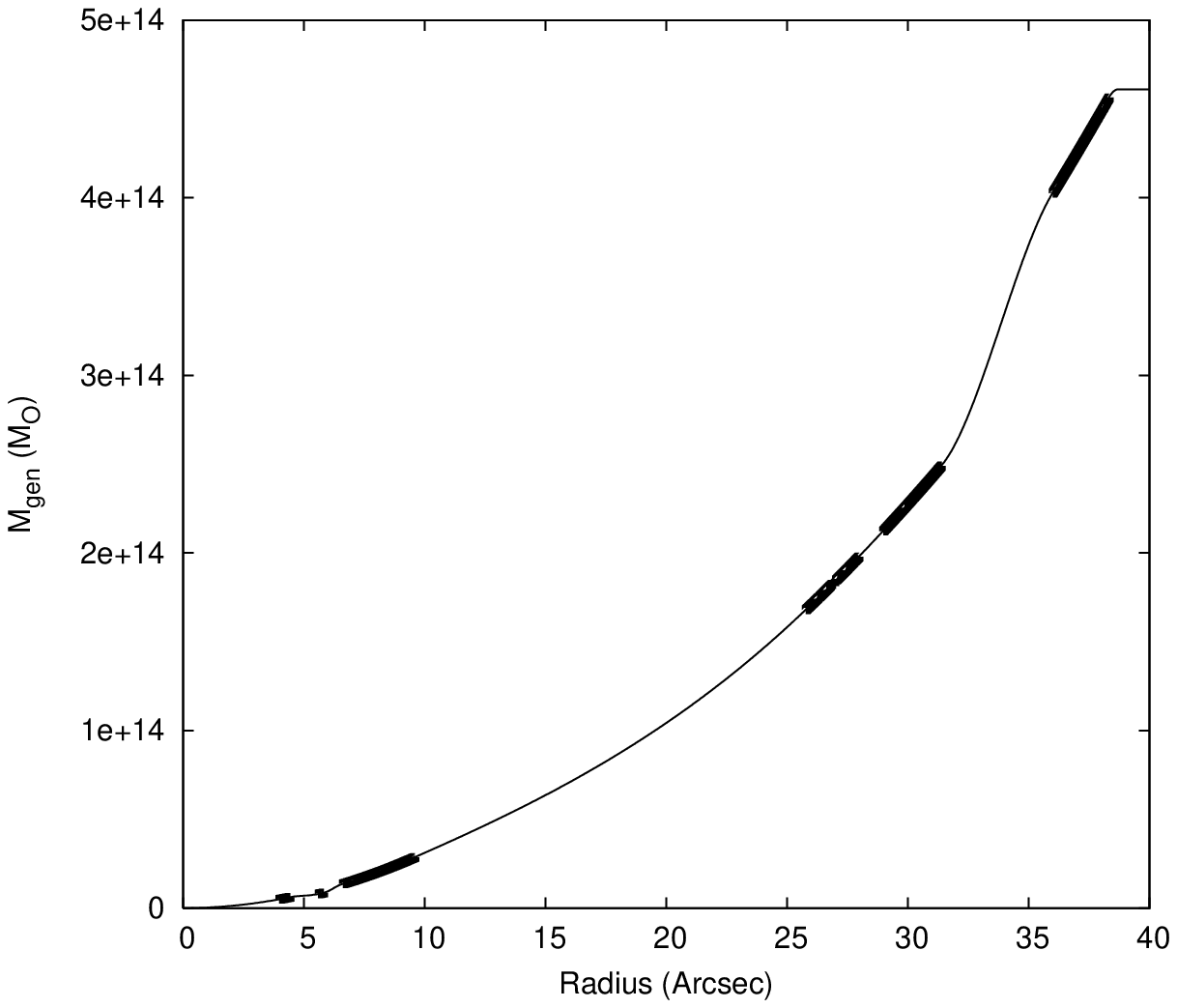}}
\qquad
\subfigure{\includegraphics[width=0.44\textwidth]{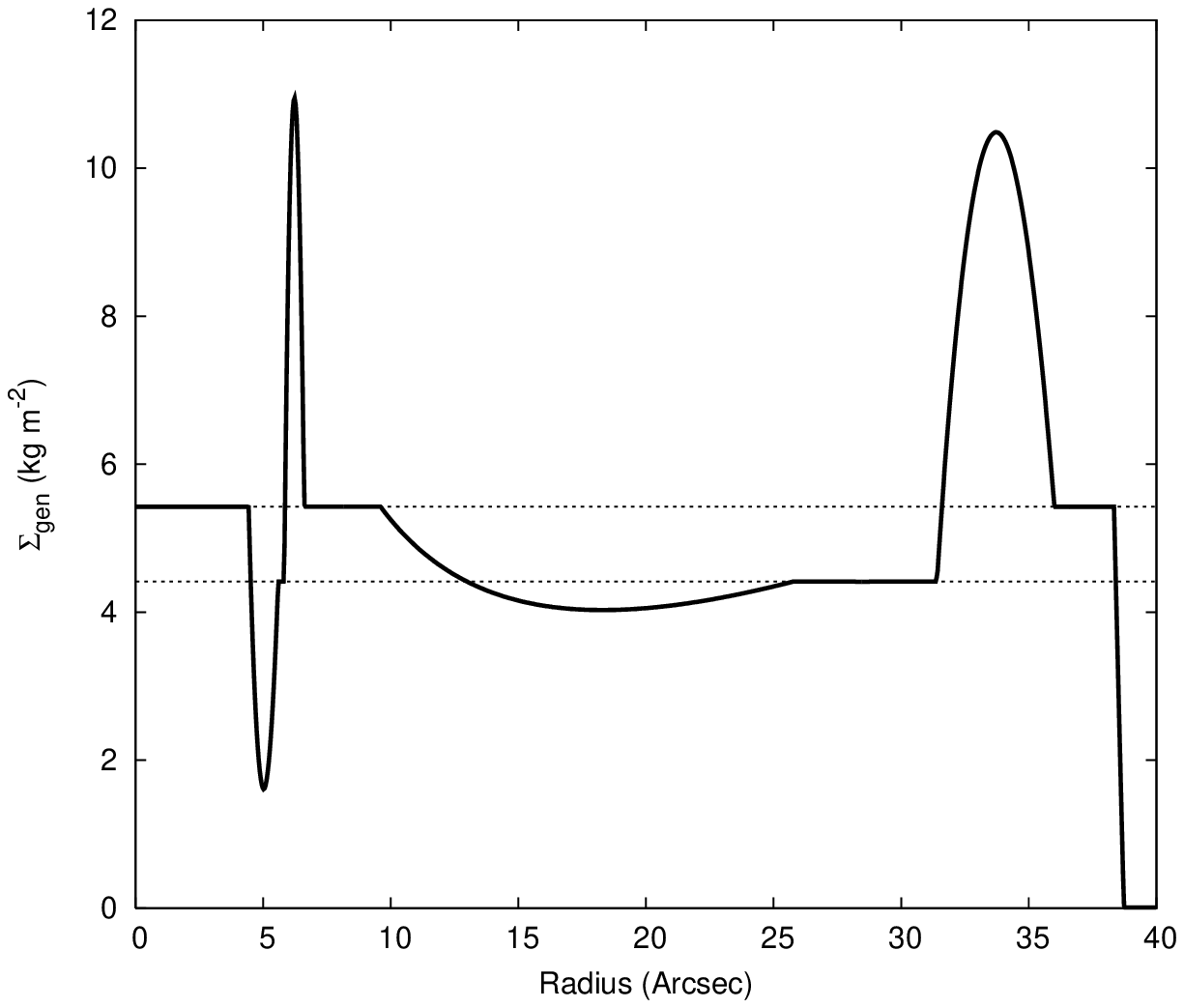}}
\caption{Left panel:~the positions of the images of each source place 
constraints on the enclosed mass $M_{\rm gen}$ (thick lines on the mass profile).
The regions in between can easily be interpolated.
Right panel:~the total mass profile in the left panel gives
rise to the density profile $\Sigma_{\rm gen}$ shown here. The dotted 
lines indicate the critical mass densities for the two sources.}
\label{fig:constraintsandprofile}
\end{figure*}

\begin{figure*}
\begin{minipage}{0.47\textwidth}
\centering
\includegraphics[height=0.29\textheight]{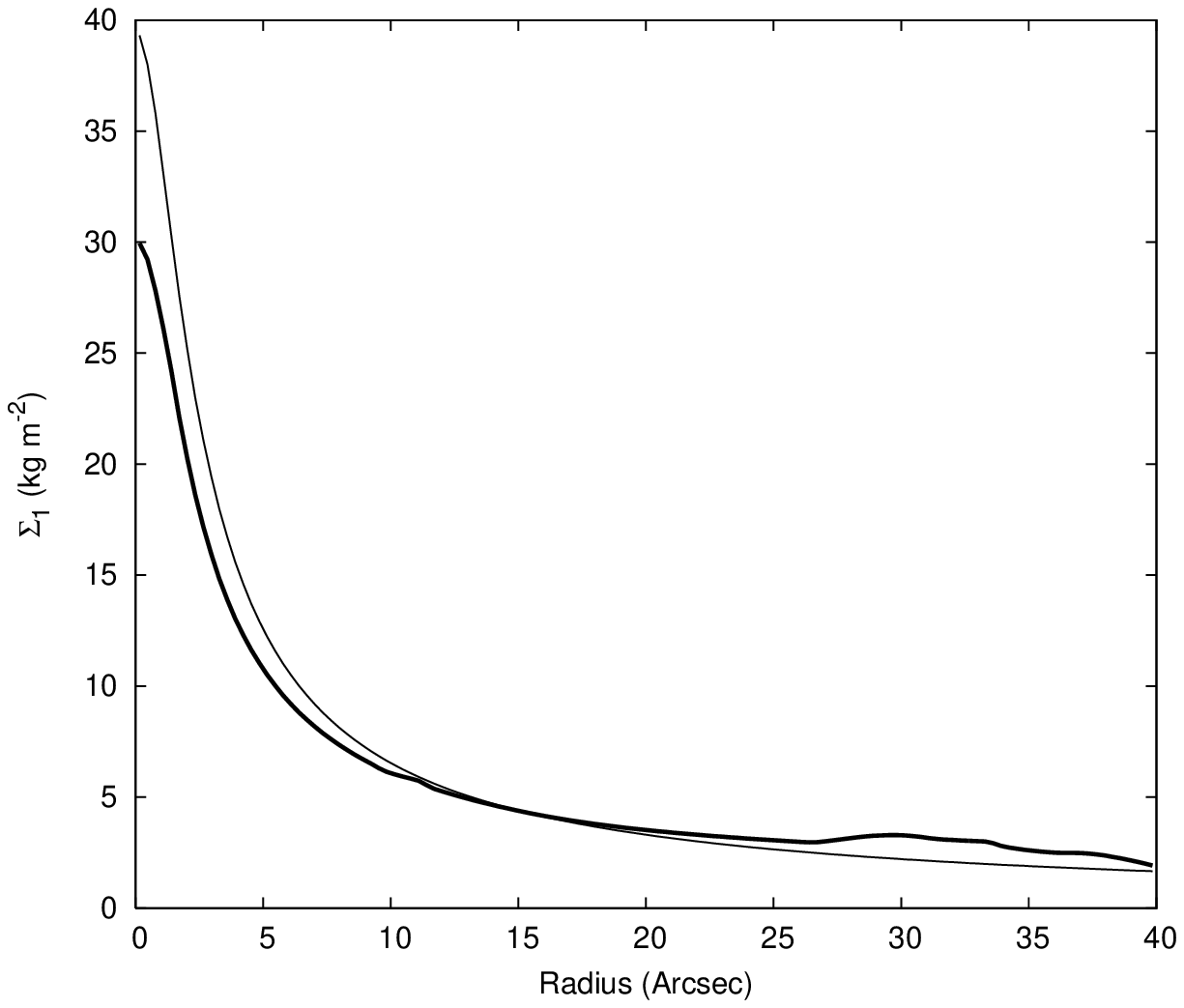}
\caption{The profile of the degenerate solution (thick line) is
compared to the profile of the original mass distribution,
a non-singular isothermal ellipse. In this example, 
$\lambda = 0.75$ was used.}
\label{fig:massprofiles}
\end{minipage}
\qquad
\begin{minipage}{0.47\textwidth}
\centering
\includegraphics[height=0.29\textheight]{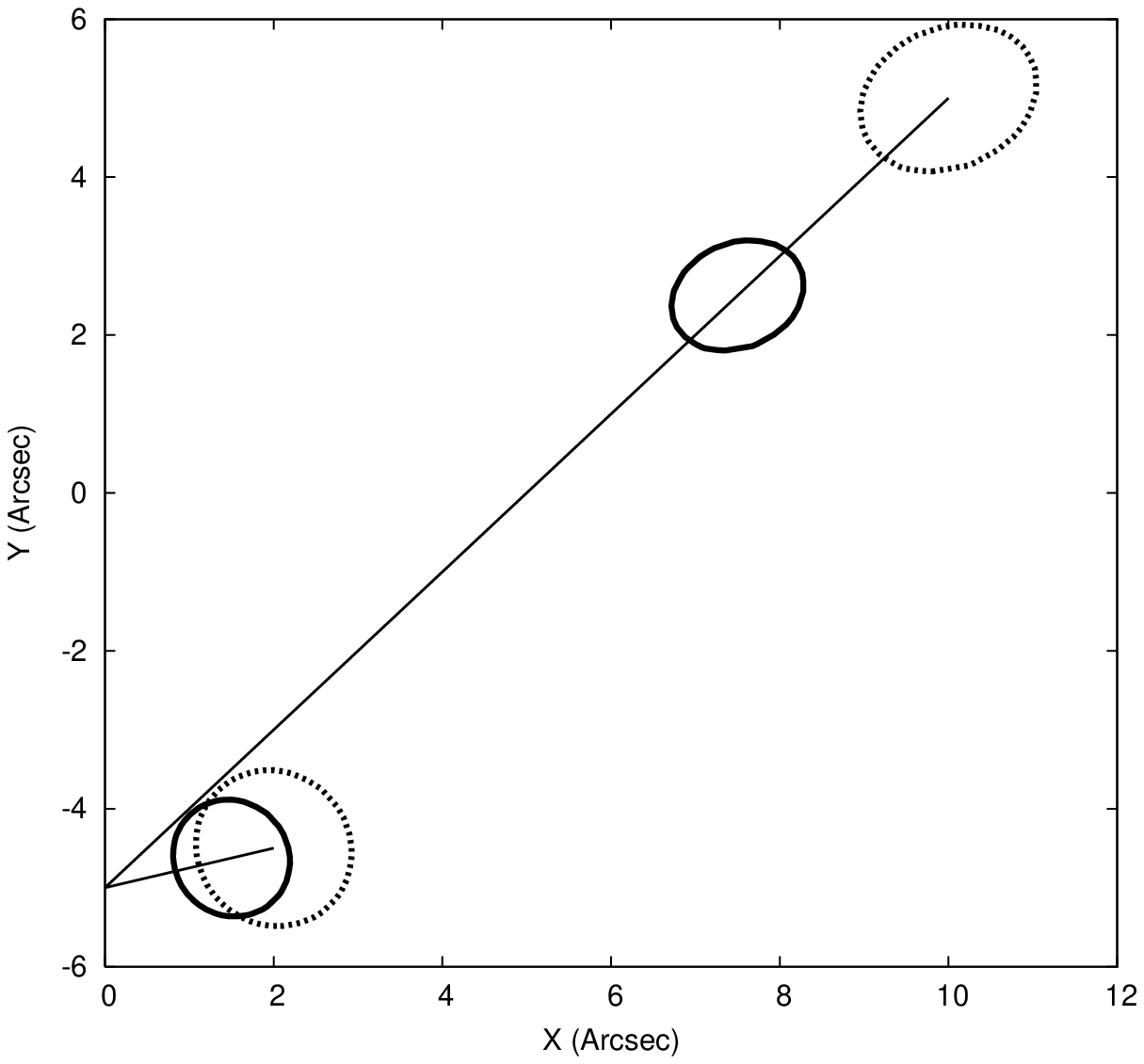}
\caption{Sources recreated by the degenerate solution (thick solid lines);
the original sources are indicated by dotted lines. The
direction of the scaling is clearly towards $(0, -5)$.}
\label{fig:sourcesizes}
\end{minipage}
\end{figure*}

\begin{figure*}
\centering
\subfigure{\includegraphics[width=0.44\textwidth]{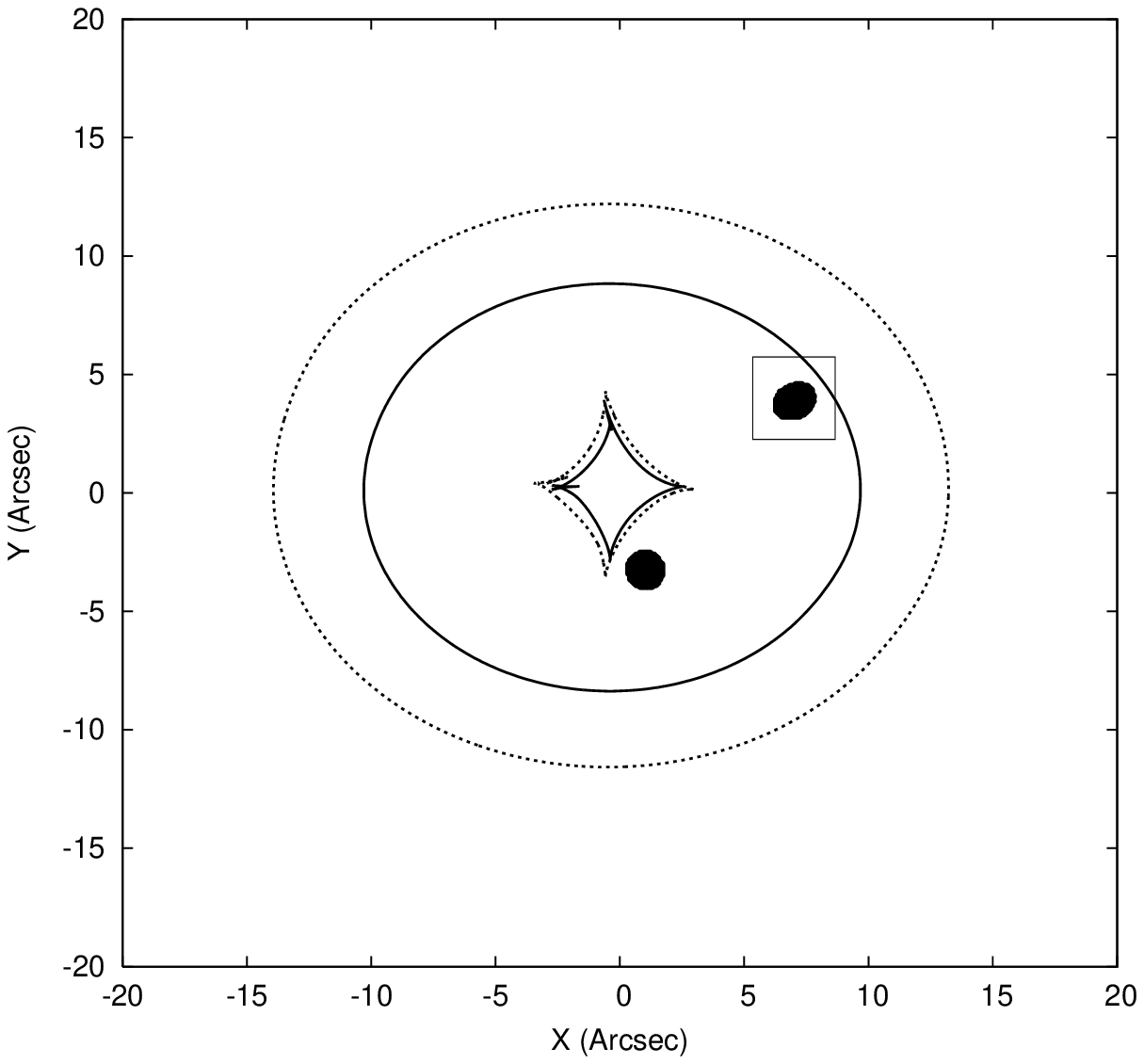}}
\qquad
\subfigure{\includegraphics[width=0.44\textwidth]{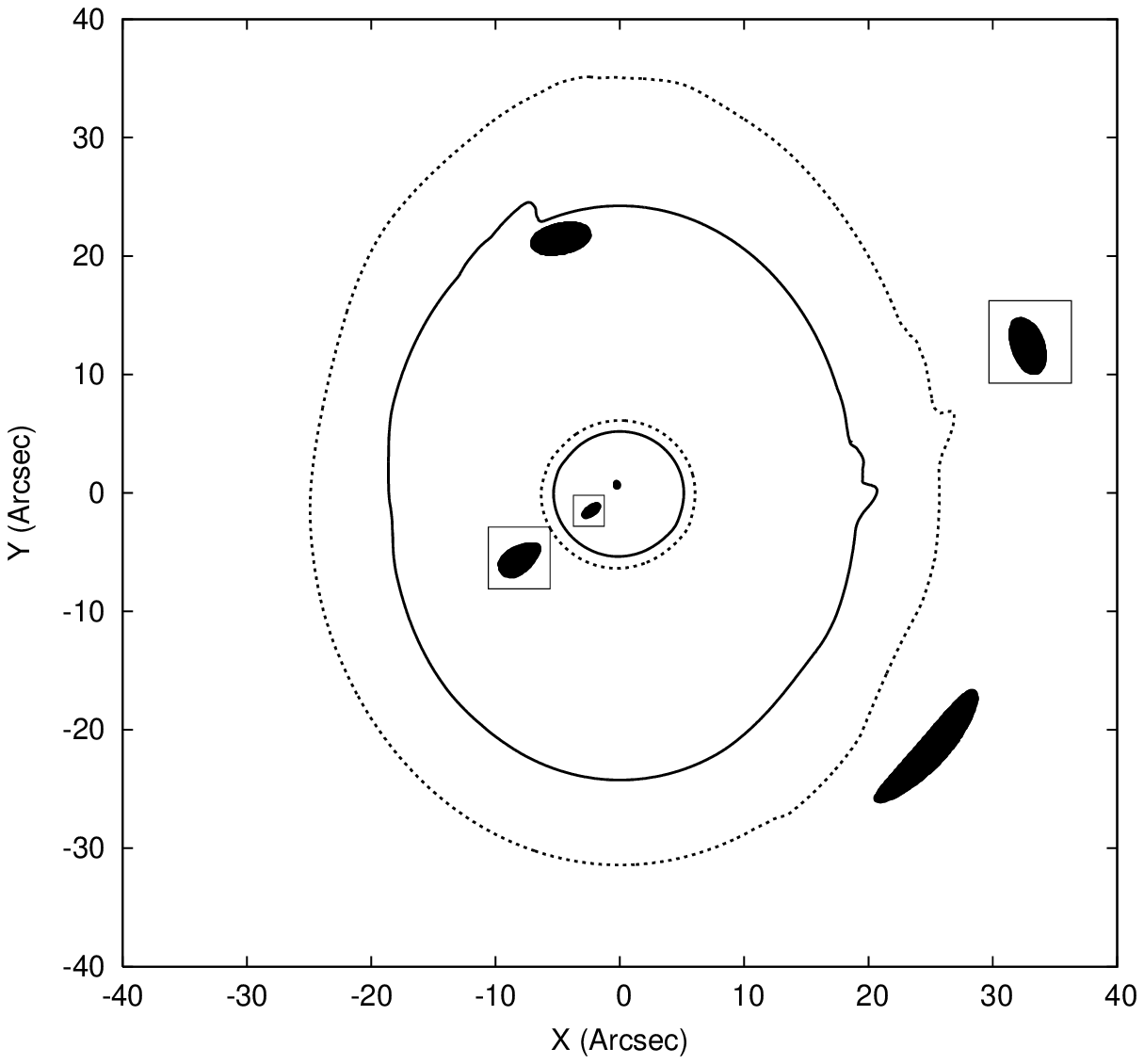}}
\caption{Left panel:~sources and caustics predicted by a degenerate solution.
Comparing with the left panel of Fig. \ref{fig:planes} one sees that
both sources and caustics are scaled versions of their original
counterparts. Right panel:~the reconstructed sources and caustics shown in 
the left panel predict the images and critical lines shown here. The same 
images as in the right panel of Fig. \ref{fig:planes} can be seen and the 
critical lines still resemble the original ones.}
\label{fig:newplanes}
\end{figure*}

\begin{figure*}
\centering
\subfigure{\includegraphics[origin=c,angle=-90,width=0.44\textwidth]{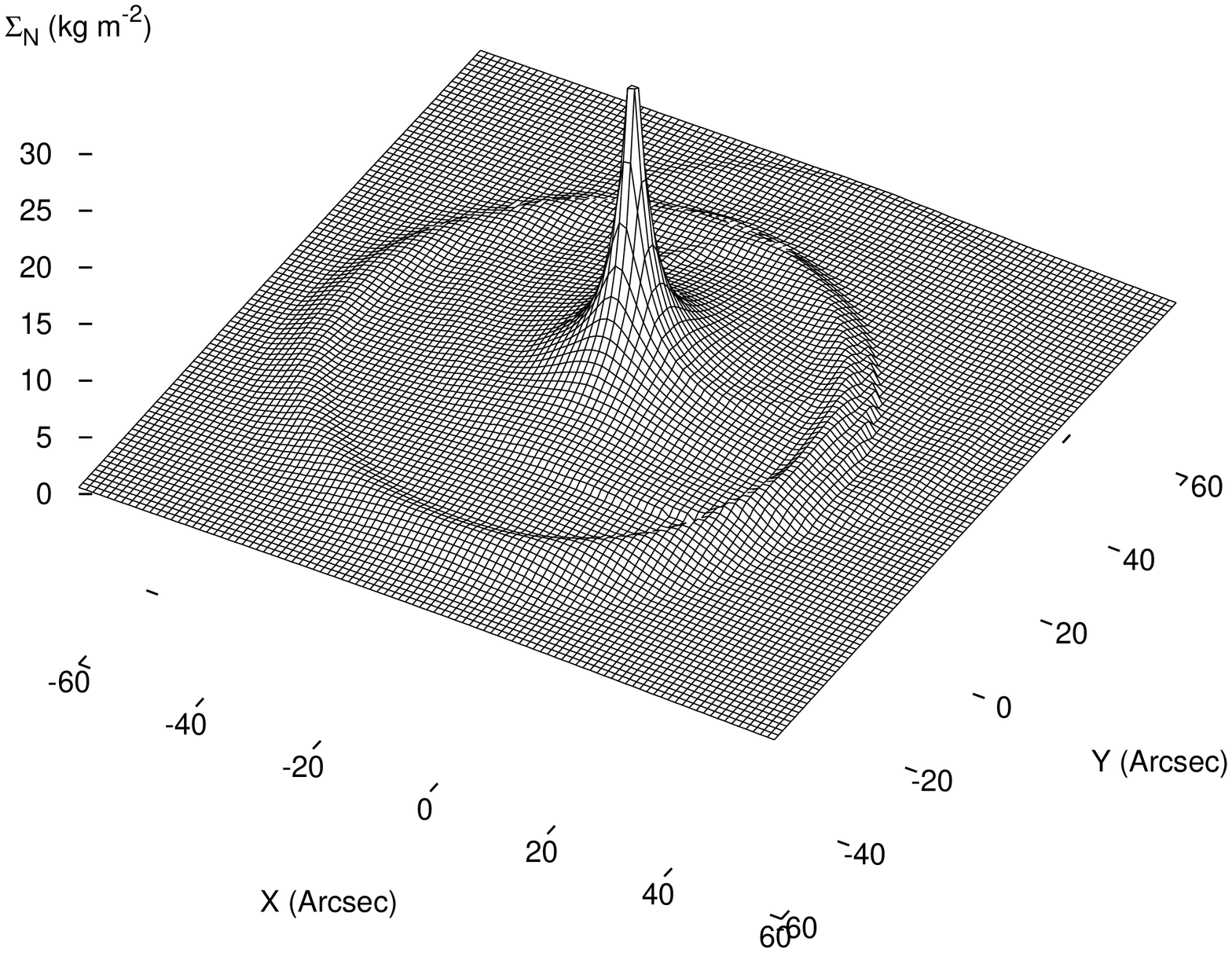}}
\qquad
\subfigure{\includegraphics[width=0.44\textwidth]{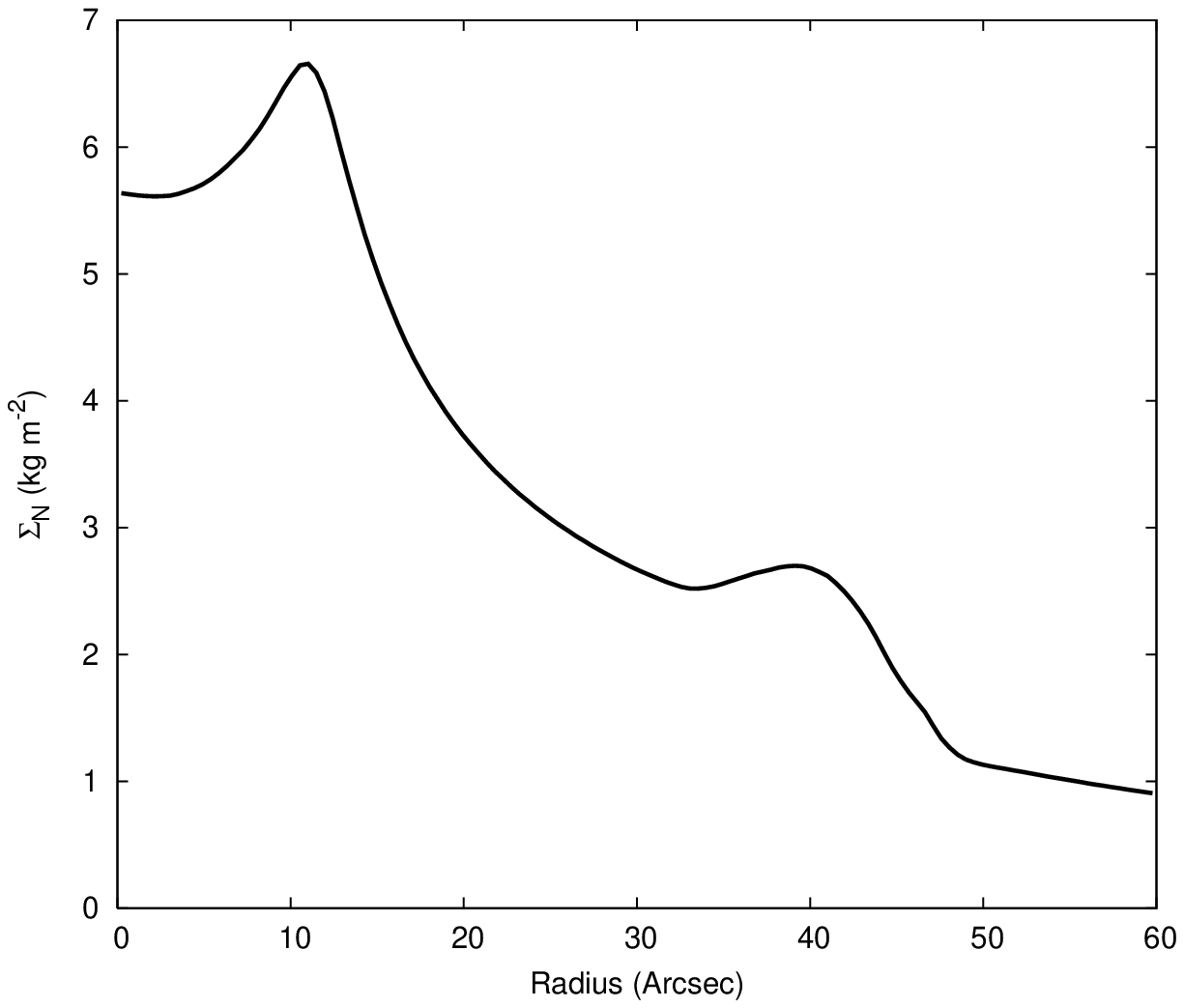}}
\caption{Left panel:~the degenerate solution which gives rise to the source 
and image planes shown in Fig. \ref{fig:newplanes}. Several ring-like 
features can be seen, the most prominent one being centered on $(-8, -8)$.
Right panel:~density profile as seen from $(-8, -8)$. Apart from the
peak of the non-singular isothermal ellipse, one can clearly see a
ring-like feature.}
\label{fig:ringstructures}
\end{figure*}

Once an appropriate center has been identified, the positions of the
images of each source can be used to calculate parts of the total mass
map $M_{\rm gen}(\theta)$, as specified by (\ref{eq:con}). In our
example, these constraints are illustrated by thick black lines
in Fig. \ref{fig:constraintsandprofile} (left panel), when using
$(0, -5)$ as the center of the distribution. The rest of the mass
map can easily be interpolated, after which the full density profile
of $\Sigma_{\rm gen}(\theta)$ can be
derived. In the left panel of Fig. \ref{fig:constraintsandprofile}, 
a third degree polynomial was used to interpolate between the 
constrained regions. The resulting density profile is plotted in 
in the right panel of the same figure and the critical densities for the
two sources are indicated with dotted lines. Note that although this
particular example does not require negative densities, in general it
is possible that this is indeed necessary. This need not be a problem,
since the resulting mass distribution will be combined with the
existing distribution $\Sigma_0(\Vec{\theta})$ (the non-singular
isothermal ellipse in this example) and may still yield an overall
positive density profile. Still, placing a positivity constraint on
the overall density profile may help to alleviate this degeneracy.

By construction, the procedure (\ref{eq:addsig}) has the same effect as
the mass-sheet degeneracy: the observed images are identical but the
reconstructed sources are scaled versions of the original ones while
the density profile of the lens has become less steep. The resulting
density profile for $\lambda = 0.75$ can be seen in
Fig. \ref{fig:massprofiles}, in which the original profile is shown as
well. Clearly, the central peak has become weaker while at larger
radii a ring of excess density has been introduced. This figure
illustrates nicely that the term steepness degeneracy still applies
to this kind of degenerate solution. When the images of
Fig. \ref{fig:planes} are projected back onto their source planes
using the new mass distribution $\Sigma_1$, the sources in
Fig. \ref{fig:sourcesizes} (solid lines) are retrieved. The fact that
the images of a specific source overlap perfectly when projected
onto the source plane proves that the constructed mass distribution
is still compatible with the observed images and can therefore 
correctly be identified as a degenerate solution. Since each
dimension is scaled by $\lambda = 0.75$, the reconstructed sources are
smaller than the original ones (dotted lines). The image also clearly
shows that the direction of the scaling is towards $(0, -5)$, the
center of the circularly symmetric $\Sigma_{\rm gen}$ which was
constructed.

Of course, since the positions of the images are not affected, one is
free to repeat the entire procedure using the newly acquired
$\Sigma_1$ as the ``original'' solution.  In general, if it is
possible to create $N$ different circularly symmetric density
distributions $\Sigma_{{\rm gen},i}$, each with another center of
symmetry $\Vec{\theta}_{{\rm c},i}$, it is easily derived that 
for any $\lambda$, the following mass distribution will still project 
the images back onto consistent sources:
\begin{equation}\label{eq:multidegen}
\Sigma_N(\Vec{\theta}) = \lambda^N \Sigma_0(\Vec{\theta}) + 
(1-\lambda)\sum_{i=1}^N \lambda^{N-i}\Sigma_{{\rm gen},i}(|\Vec{\theta}-\Vec{\theta}_{{\rm c},i}|) \mpt
\end{equation}
This way, the mass distribution which is added to $\Sigma_0$ need not
possess circular symmetry anymore and its density profile can become
much more complex. Equation (\ref{eq:multidegen}) is important from
a practical point of view: the target scale $\lambda^N$ can easily
be reached by using $N$ generators, each producing only a very small 
effect. If a large number of suitable center positions can be found,
this can severely reduce the amount of substructure introduced by the
procedure.

An example of degenerate source and image planes
obtained by using $N = 100$ different $\Sigma_{{\rm gen},i}$ can be
seen in Fig. \ref{fig:newplanes}. Each source is scaled by a factor
$\lambda^N = 0.75$ in each dimension; the caustics are scaled as
well. The critical lines still show the same general structure.  The
mass distribution of the degenerate solution can be seen in Fig.
\ref{fig:ringstructures} (left panel) and contains several ring-like
structures, the most prominent one being centered on $(-8, -8)$. This
can also be clearly seen when the density profile is calculated using
$(-8, -8)$ as the center (right panel of
Fig. \ref{fig:ringstructures}). The first peak in this plot is due to
the non-singular isothermal ellipse, the second one is caused by the
ring-like substructure of the degenerate solution. Note that the ring
is not caused by one particular generator, but is a combined effect.

As an aside, even when it is impossible to rescale the sources, it may
still be possible to introduce ring-shaped features. One only needs a
ring-shaped region without data points. A circularly symmetric
distribution which is zero everywhere but which fluctuates in the
ring-shaped region in such a way that the total mass inside the region
is zero as well, can simply be added to the original, again thanks to
equation (\ref{eq:alphasymm}). Doing so can obviously introduce a
ring-like feature in the mass distribution of the lens.

\section{Discussion and conclusion}\label{sec:conc}

In this article we have presented a straightforward extension of
the mass-sheet degeneracy to multiple redshifts. Although there
is no actual sheet of mass involved, the procedure and effect of
the degeneracy are the same. In both cases, the existing mass
distribution is rescaled and a component is added, leading
to a rescaling of the sources. This justifies placing the
degeneracy described above in the same category as the mass-sheet
or steepness degeneracy. Although we illustrated the construction of 
degenerate solutions using a strong lensing example, it is clear 
that weak lensing studies can be affected as well. In particular, 
if weak lensing data alone leads to solutions prone to the 
mass-sheet degeneracy, it will not suffice to add strong 
lensing data of a single source to break the degeneracy if 
strong and weak lensing regions do not overlap. Otherwise, a 
similar construction as above is possible, using one critical 
density in the weak lensing region and another in the strong 
lensing region. 

However, as was shown by \citet{Bradac2004}, using weak lensing 
data it is in principle possible to break the mass-sheet 
degeneracy -- including the extension described here -- if 
individual source redshifts are available and if sources with a 
rather high distortion are included. Another way to break the
degeneracy is to add information about the magnification, for
example by using source number statistics \citep{Broadhurst1995}
or Type Ia supernovae observations \citep{Holz}. Additional
information about stellar dynamics in the gravitational lens
can also help to break the degeneracy \citep{Koopmans2006}.
Since the mass-sheet degeneracy rescales the time delay
surface, time delay measurements can be used to break it as well.

This degeneracy seems to explain what we observed during tests of our
non-parametric inversion algorithm \citep{Liesenborgs2}. Although an
example using two sources was used to avoid the mass-sheet degeneracy,
we found that our algorithm did not succeed in finding the correct
source sizes. However, the general shape and features of the 
reconstructed mass map did resemble closely the true mass distribution,
which was used to create the images that served as input for the inversion
routine. It is now clear that images of two sources at different redshifts
do not provide enough constraints to firmly establish the scale
of these sources.

The procedure outlined above still allows much freedom in the
interpolation scheme, and therefore in the precise shape of 
$\Sigma_{\rm gen}$. It is clear, however, that in general it will
always be necessary to introduce substructure in $\Sigma_{\rm gen}$,
which may lead to the presence of substructure in the new
mass map. The presence of such substructure will limit the
range in which the parameter $\lambda$ may lie: if too
much substructure is introduced, additional images will be
predicted and the created mass distribution will no
longer be compatible with the observed situation. However, by 
using several versions of $\Sigma_{\rm gen}$, each with a different 
center of symmetry, the amount of substructure that needs to
be introduced to obtain a specific scaling can be limited.

It is straightforward to extend the procedure to more than two
sources, but at some point it will no longer be possible to
construct such degenerate solutions. How many sources are
needed will depend on a number of factors. The number of the 
images and their sizes play an important role, as this 
determines how easily their annuli will overlap and therefore
how difficult it will be to find appropriate scaling centers.
The precise redshifts of the sources will determine how much 
substructure needs to be introduced, since the redshifts will
determine if the corresponding critical densities lie close
to each other or differ a lot. If the latter is the case,
automatically more substructure will need to be introduced,
allowing only a limited range of $\lambda$ values to avoid
the prediction of extra images.

Although at a first glance the mass-sheet or steepness degeneracy 
is easily broken, this article shows that it is in fact a lot 
more difficult to do so and a relatively large number of 
constraints may be needed. The examples above also suggest
that circularly symmetric features should be distrusted, as they 
are easily introduced in degenerate solutions.

\section*{Acknowledgment}

We thank our referee, Prasenjit Saha, for his comments on this
article.

\bsp 
\label{lastpage}

\end{document}